\documentclass[%
 reprint,
 amsmath,amssymb,
 aps,
]{revtex4-2}

\usepackage{graphicx}
\usepackage{dcolumn}
\usepackage{bm}
\usepackage{xcolor}

\newcommand{\Dq}{\Delta_c}
\newcommand{\JH}{J_H}

\begin{document}

\preprint{APS/123-QED}


\title{Non-Kitaev vs. Kitaev Honeycomb Cobaltates}
\author{Xiaoyu Liu$^1$}
\author{Hae-Young Kee$^{1,2}$}
\email[]{hykee@physics.utoronto.ca}
\affiliation{Department of Physics, University of Toronto, Ontario, Canada M5S 1A7}
\affiliation{Canadian Institute for Advanced Research, CIFAR Program in Quantum Materials, Toronto, Ontario, Canada M5G 1M1 }
\date{\today}

\begin{abstract}
Recently, honeycomb cobaltates with 3$d^7$ were proposed to display Kitaev physics despite weak spin-orbit coupling.
However, other theoretical and experimental works found leading XXZ Heisenberg and negligible Kitaev interactions in 
BaCo$_2$(AsO$_4$)$_2$ (BCAO), 
which calls for a further study to clarify the origin of the discrepancies. 
Here we derive the analytical expressions of the spin model using strong-coupling perturbation theory.  
With tight binding parameters obtained by {\it ab-initio} calculations for idealized honeycomb BCAO, we find that the largest intraorbital $t_{2g}-t_{2g}$ exchange path, which was assumed to be small in the earlier theory proposal, leads to a ferromagnetic (FM) Heisenberg interaction. This becomes the dominant interaction, as other
$t_{2g}-e_g$ and $e_g-e_g$
contributions almost cancel each other. 
Exactly the same assumed-to-be-small channel also generates an antiferromagnetic Kitaev interaction, which then cancels a FM Kitaev interaction from $t_{2g}-e_g$ paths, resulting in a small Kitaev interaction. 
Under the trigonal distortion, the preeminent isotropic Heisenberg becomes an anisotropic XXZ model as expected, which is the case for BCAO.
However, in Na$_3$Co$_2$SbO$_6$ the intraorbital $t_{2g}-t_{2g}$ hopping is smaller and comparable to the $t_{2g}-e_g$ hopping, leading to a delicate competition between enhanced Kitaev and reduced Heisenberg interactions.

\end{abstract}

\maketitle

\section{Introduction}
The Kitaev model which consists of bond-dependent Ising interaction on the honeycomb lattice is exactly solvable, and its ground state is the Kitaev spin liquid (KSL) with unusual excitations.\cite{kitaev_2006}
Due to the bond-dependent interaction character, its materialization requires angular momentum and spin degrees of freedom and their couplings, i.e., spin-orbit coupling (SOC).\cite{jackeli_mott_2009} There has been a surge of studies on Kitaev candidate materials which may exhibit the KSL.
The first proposed candidate was a group of $5d^5$ iridium oxides \cite{jackeli_mott_2009,singh_afm_2010,singh_Li_2012}. Later, $\alpha$-RuCl$_3$\cite{plumb_RuCl3_2014,Sandilands2015PRL,Kim2015PRB,Banerjee2016NM} with $4d^5$ was suggested to host the dominant Kitaev interaction despite the reduced SOC compared with iridates.\cite{Kim2015PRB}
The single hole on $t_{2g}$ orbitals has spin S=1/2 and pseudo angular momentum L$_{\rm eff}$=1, giving rise to the $J_{\rm eff}=1/2$ doublet under strong SOC. The indirect exchange processes of $J_{\rm eff}=1/2$ via $p$-orbitals lead to the Kitaev interaction.\cite{jackeli_mott_2009}

However these candidate materials are magnetically ordered at low temperatures.\cite{singh_afm_2010,singh_Li_2012,liu_ordered_2011,ye_zigzag_2012,sesars_ordered_2015,johnson_ordered_2015,cao_ordered_2016,takagi_concept_2019} This suggests that additional exchange interactions other than the Kitaev interaction are present. In particular, Heisenberg interaction arising from direct hopping between $d$ orbitals is non-negligible.\cite{chaloupka_kh_2010,chaloupka_zigzag_2013}
In addition, another bond-dependent interaction known as $\Gamma$ interaction\cite{rau_jkg_2014} is significant in $4d$ and $5d$ materials. In real materials, further interactions are present. 
For example, other bond-dependent interaction called $\Gamma^\prime$ is introduced due to the trigonal distortion.\cite{rau2014arXiv} 

Thus a battle among the symmetry-allowed Kitaev ($K$), Heisenberg ($J$),  $\Gamma$, and $\Gamma^\prime$ interactions selects Kitaev materials referring to materials with the dominant Kitaev interaction. 
To build the Kitaev dominant systems, 
 Liu {\it et al}~\cite{huimei_prb_2018} and Sano {\it et al}~\cite{sano_KHd7_2018} proposed to use a less extended $3d$ orbitals,
 as it may involve a smaller $d-d$ hopping integral.
They suggested $3d^7$ cobaltates would be good  candidates despite small SOC in $3d$ systems. 
The $3d^7$ has one hole in $t_{2g}$ ($t_{2g}^5$) and two holes in $e_g$ ($e_g^2$) leading to a total spin S=3/2 and total angular momentum L=1. With SOC, the lowest state is $J_{\rm eff} =1/2$. 
It was shown that the $t_{2g}-e_g$ channels generate large AFM $J$ and FM $K$ interactions. After taking the $e_g-e_g$ exchange contribution, the cancellation of Heisenberg interaction occurs, which makes the Kitaev interaction dominant. The $t_{2g}-t_{2g}$ contributions to $J$ and $K$ were negligible, see Fig.~2 in Ref. \cite{huimei_prl_2020}.

Motivated by the proposal, several  theoretical and experimental works have been carried out on various rhombohedral cobaltates. \cite{Songvilay2020PRB,Yao2020PRB,Vivanco2020PRB,Zhang2021arXiv,Kim2021,Chen2021PRB,Lee2021PRB,Samarakoon2021PRB,Sanders2022PRB,Kim2022JPCM,Yang2022PRB}
Theoretical studies found that BaCo$_2$(AsO$_4$)$_2$ (BCAO) is better described by the XXZ model with significant third nearest neighbour (n.n.) Heisenberg ($J_3$) but negligible Kitaev interactions\cite{das_XY_2021,maksimov_ab_2022,winter_d7_2022,halloran_geometrical_2022} similar to
an earlier study on  BaCo$_2$(PO$_4$)$_2$ (BCPO)\cite{Nair2018PRB}.
More recently, a combined experimental and theoretical work\cite{halloran_geometrical_2022} compared the two scenarios, XXZ-$J$-$J_3$ and $JK\Gamma\Gamma'$ models in detail for BCAO. Using these models, they fitted high-field magnon dispersion obtained by inelastic neutron scattering measurement and showed that the former model fits the experimental data well rather than the latter. 
These works together question if cobaltates fall into the Kitaev candidates and call for a closer inspection on how the exchange processes add up or cancel out each other. 

In this work we investigate the exchange processes for $3d^7$ using the strong-coupling perturbation theory and {\it ab-initio} calculations to address the origin of the discrepancies among the previous works in BCAO. Starting from the ideal honeycomb lattice, we
find that an {\it intraorbital} hopping channel (denoted by $t_3$) among the $t_{2g}-t_{2g}$ exchange paths, assumed to be negligible in Ref. \cite{huimei_prb_2018,sano_KHd7_2018}, is significant. Taking into account this exchange path together with Hund's coupling, the FM Heisenberg interaction is greatly boosted. This becomes the major Heisenberg interaction, because other contributions from the $t_{2g}-e_g$ and $e_g-e_g$ paths almost cancel each other. Exactly the same assumed-to-be-small channel also generates the AFM Kitaev interaction which then cancels the proposed FM Kitaev interaction from the $t_{2g}-e_g$ path, resulting in a small Kitaev interaction. 
Under the trigonal distortion, the isotropic Heisenberg interaction becomes an anisotropic XXZ  interaction which dominates over other interactions.
On the other hand, Na$_3$Co$_2$SbO$_6$ (NCSO), the $t_{2g}-e_g$ hopping
becomes comparable to $t_3$ as $t_3$ decreases\cite{Kim2022JPCM}, which will reduce the Heisenberg interaction, but
boost the FM Kitaev interaction.

The rest of the paper is organized as follows. In Sec.~II we review the onsite Hamiltonian of Co honeycomb. In Sec.~III we introduce the various direct and indirect hoppings considered in our model. Sec.~IV introduces the derivation of spin Hamiltonian and gives analytical expressions for various exchange interactions. We present {\it ab initio} parameters and results for BCAO as an example in Sec.~V. The effect of the trigonal distortion is discussed in Sec.~VI and finally the discussion and conclusion in Sec. ~VII.

\section{onsite hamiltonian}
\begin{figure}
    \centering
    \includegraphics[scale=1]{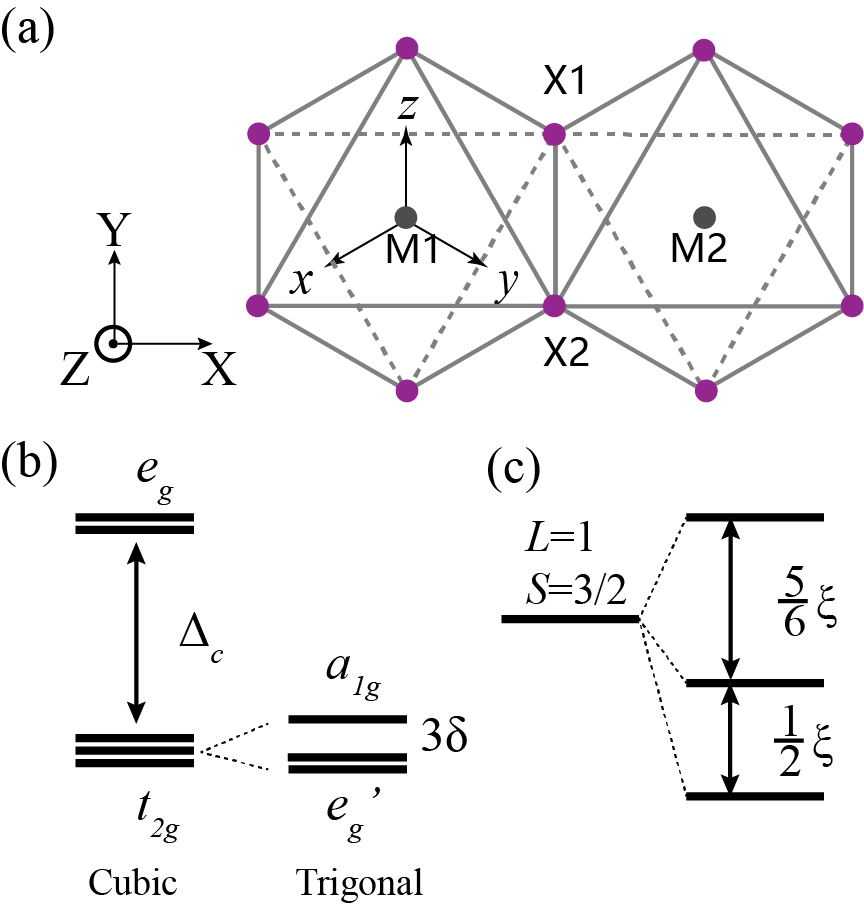}
    \caption{(a) Crystal structure for edge-sharing honeycomb octahedral lattice. The local octahedral coordinates ($xyz$) and the global basis ($XYZ$) are shown. (b) Splittings of $3d$ orbitals under cubic ($\Delta_c$) and trigonal ($\delta$) crystal fields. (c) Energy level splittings of $L=1, S=\frac{3}{2}$ states under SOC ($\xi$).}
    \label{fig:illustration}
\end{figure}

The structure of honeycomb cobaltates is similar to other edge-sharing octahedral honeycomb materials (see Fig.~\ref{fig:illustration}(a)). For these materials, the onsite Hamiltonian is expressed by
\begin{equation}\label{eq:htot}
H_{\rm{tot}}=H_{\rm{Coulomb}}+H_{\rm{cubic}}+H_{\rm{SOC}}+H_{\rm{trig}}.
\end{equation}
For simplicity, the isotropic Kanamori interaction\cite{kanamori_1963}
is used which is a good approximation to the full Coulomb interaction with 3- and 4-orbital effects\cite{sugano_multiplets_2014,wang_edrixs_2019,coury_hubbard_2016,xiaoyu_SIA_2022}. 
Within this approximation, the interaction part $H_{\rm{Coulomb}}$ is written as
\begin{equation}
\begin{aligned}\label{eq:kanamori}
H_{\rm{Coulomb}} &= U\sum_\alpha n_{\alpha\uparrow} n_{\alpha\downarrow} +
\frac{U'}{2} \sum_{\alpha\neq\beta,\sigma,\sigma'} n_{\alpha \sigma} n_{\beta \sigma'}\\
&-\frac{\JH}{2}\sum_{\alpha\neq\beta,\sigma\sigma'}c_{\alpha\sigma}^\dagger c_{\beta\sigma'}^\dagger c_{\beta \sigma} c_{\alpha \sigma'}\\
&+ \JH\sum_{\alpha\neq\beta} c_{\alpha\uparrow}^\dagger c_{\alpha\downarrow}^\dagger c_{\beta \downarrow} c_{\beta \uparrow},
\end{aligned}
\end{equation}
where $U$ and $U'$ are intra- and interorbital Coulomb interactions respectively, and $\JH$ is the Hund's coupling. 

There are two kinds of crystal field splittings (CFS) in edge-sharing octahedral honeycomb lattices. One is the cubic CFS due to the octahedral cage formed by oxygen atoms, see Fig.~\ref{fig:illustration}(b). It is usually around 1~eV and is comparable to the Hund's coupling $\JH$ for $3d$ orbitals. The other is the trigonal CFS induced by the trigonal distortion of the 2D honeycomb lattice (compress or elongate of octahedral cages along $Z$ direction shown in Fig.~\ref{fig:illustration}(a)). This is usually small (from several meV to several tens of meV). 
Combining the two crystal field effects, we have
\begin{equation}\label{eq:cf}
    H_{CFS}=\left(
\begin{array}{ccccc}
    \Dq & 0 & 0 & 0 & 0  \\
    0 & \Dq & 0 & 0 & 0\\
    0 & 0 & 0 & \delta & \delta\\
    0 & 0 & \delta & 0 & \delta\\
    0 & 0 & \delta & \delta & 0 
    \end{array}
\right),
\end{equation}
written in the basis of $d_{x^2-^2}, d_{3z^2-r^2}, d_{yz}, d_{xz}, d_{xy}$. 
$\Dq$ is the splitting between $t_{2g}$ and $e_g$ orbitals by the cubic CFS and $\delta$ is the trigonal distortion. 

Due to the large Hund's coupling, $3d^7$ forms high spin state with total angular momentum $L=1$ and total spin $S=3/2$. Under SOC $\lambda L\cdot S$, the lowest states form a $J_{\rm eff}=\frac{1}{2}$ doublet (see Fig.~\ref{fig:illustration}(b)) with wavefunction 
\begin{equation}
\begin{aligned}
|+\tilde{\frac{1}{2}}\rangle &= \frac{1}{\sqrt{2}}|\frac{3}{2},-1\rangle-\frac{1}{\sqrt{3}}|\frac{1}{2},0\rangle+\frac{1}{\sqrt{6}}|-\frac{1}{2},1\rangle,\\
|-\tilde{\frac{1}{2}}\rangle &= \frac{1}{\sqrt{2}}|-\frac{3}{2},1\rangle-\frac{1}{\sqrt{3}}|-\frac{1}{2},0\rangle+\frac{1}{\sqrt{6}}|\frac{1}{2},-1\rangle.\\
\end{aligned}
\end{equation}
They are written in $|m_S, m_L\rangle$ basis with $m_S$ being the magnetic spin moment and $m_L$ magnetic orbital angular moment \cite{huimei_prb_2018,huimei_prl_2020,sano_KHd7_2018,liu2021IJMP,winter_d7_2022}. 
In this work, We use $H_{\rm{SOC}} = \xi\sum_i \bf{l}_i \cdot \bf{s}_i$ 
where $\xi$ is the atomic SOC strength for Co atoms. The relation to the other SOC scheme, $\lambda \sum_i {\bf L}_i \cdot {\bf S}_i$ is that $\lambda=\xi/3$ as there are three holes. The SOC-induced energy splittings are shown in Fig.~\ref{fig:illustration}(c).

\section{Exchange paths}
Having set the onsite Hamiltonian, we now consider the hopping paths to determine exchange processes. 
As shown in  Fig.~\ref{fig:illustration}(a), bond $M_1-M_2$ has $C_{2v}$ local symmetry and the symmetry allowed hoppings for the ideal honeycomb lattice are
\begin{equation}\label{eq:d-d}
    T_{dd}=\left(
\begin{array}{ccccc}
    t_5 & 0 & 0 & 0 & 0  \\
    0 & t_4 & 0 & 0 & t_6\\
    0 & 0 & t_1 & t_2 & 0\\
    0 & 0 & t_2 & t_1 & 0\\
    0 & t_6 & 0 & 0 & t_3 
    \end{array}
\right).
\end{equation}
Here $t_1, t_3, t_4$ and $t_5$ are intraorbital direct hoppings between $d_{yz/xz}$, $d_{xy}$, $d_{3z^2-r^2}$ and $d_{x^2-y^2}$ respectively. $t_2$ is the hopping between $d_{yz}$ and $d_{xz}$ which includes both direct and indirect hoppings.  $t_6$ is the hopping between $t_{2g}$ and $e_g$ manifolds and also includes both direct and indirect hoppings. Hoppings of other bonds are related to $z$-bond by $C_3$ symmetry. 
The symmetry allowed $p-d$ hopping is parameterized as (take bond $M_1-X_1$ as an example, see Fig.~\ref{fig:illustration}(a))
\begin{equation}
    T_{dp}=\left(
\begin{array}{ccc}
    \frac{\sqrt{3}}{2} t_{pd\sigma} & 0 & 0   \\
    -\frac{1}{2} t_{pd\sigma} & 0 & 0 \\
    0 & 0 & 0 \\
    0 & 0 & t_{pd\pi} \\
    0 & t_{pd\pi} & 0 
    \end{array}
\right).
\end{equation}
These hopping parameters could be obtained from DFT calculations once a target system is chosen as will be discussed in Sec.~V. 

\section{Spin model}
With both the onsite and intersite Hamiltonians, we build the spin model using the strong-coupling perturbation theory by projecting ground states doublets. 
For the edge-sharing ideal honeycomb octahedral lattice, the generic nearest neighbor (n.n.) spin model is given by\cite{rau_jkg_2014}
\begin{equation}\label{eq:spin_model}
\begin{aligned}
H_{\rm{spin}} = & \sum_{\langle i j \rangle\in \alpha\beta(\gamma)}
J\textbf{S}_i\cdot\textbf{S}_j+
K S_i^\gamma S_j^\gamma+
\Gamma (S_i^\alpha S_j^\beta+S_i^\beta S_j^\alpha).\\
\end{aligned}
\end{equation}
Here the $\alpha,\beta,\gamma$ are the local octahedral coordinates $x,y,z$. 
$J$,$K$ and $\Gamma$ are the isotropic Heisenberg, the bond-dependent Kitaev and the bond-dependent off-diagonal terms respectively. Other bond-dependent off-diagonal terms are forbidden in the ideal circumstances.  
To determine the relative strength among them, we present three different types of exchange processes and show which combinations determine the major interaction.

\subsection{Inter-site $U$ process}
A virtual hopping of one $d$ electron between neighbouring Co atoms through second-order process ( $d^7d^7-d^6d^8-d^7d^7$ or $d^7d^7-d^8d^6-d^7d^7$)  can lower the total energy and thus contributes to the spin Hamiltonian. 
The analytical expressions of spin interactions from all second-order processes are listed in Table~\ref{table:2nd} in Appendix B.

The major difference from our finding and conclusion from Ref. \cite{huimei_prb_2018,huimei_prl_2020}
is the strength of $t_3$ hopping integral. 
In Ref. \cite{huimei_prb_2018,huimei_prl_2020}, it was assumed that 
$t_3$ ($t^\prime$ in their notation) is negligible based on less-extended 3d orbitals. However, since 3d systems have a smaller lattice constant, we expect a large $t_3$ compared to other hopping integrals.
Indeed the large $t_3$ hopping integral was reported in several honeycomb cobaltates in Ref. \cite{das_XY_2021,winter_d7_2022}.

\begin{figure}
    \centering
    \includegraphics[width=0.4\columnwidth]{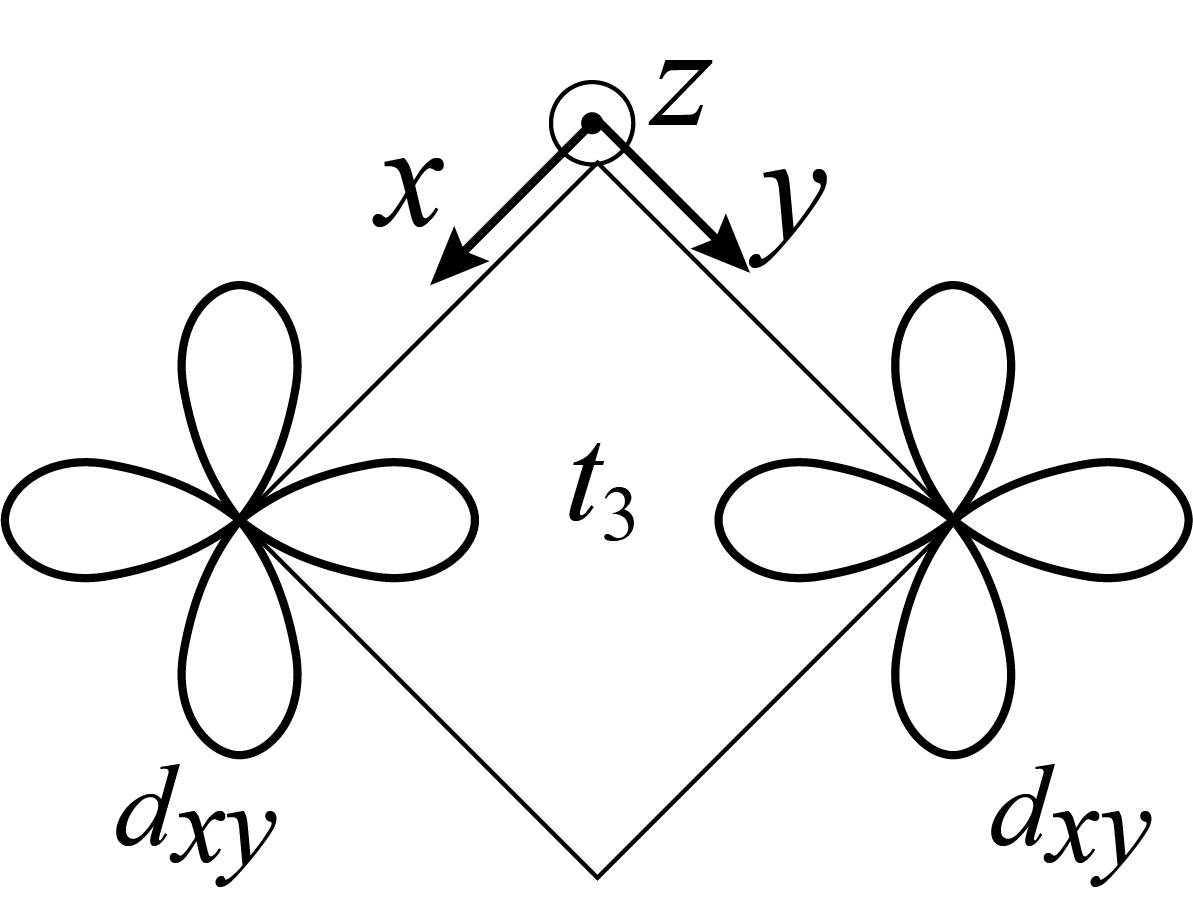}
    \caption{The dominant second order inter-site $U$ process is the direct hopping between neighbouring $d_{xy}$ orbitals. }
    \label{fig:2nd}
\end{figure}

The $t_3$ associated exchange process shown in Fig.~\ref{fig:2nd} can be summarized by the following Heisenberg and Kitaev interactions:
\begin{equation}
\begin{aligned}
    J &= \dfrac{2}{243}\left(-\dfrac{27}{U-3\JH} + \dfrac{43}{U+\JH} + \dfrac{8}{U+4\JH}\right) t_3^2,\\
    K &= \dfrac{2}{81}\left(\dfrac{3}{U-3\JH} - \dfrac{7}{U+\JH} - \dfrac{2}{U+4\JH}\right) t_3^2.
\end{aligned}
\label{JK}
\end{equation}
The Hund's coupling enters into these expressions explicitly due to the energy differences between intermediate state $d^6d^8$ (or $d^8d^6$) and the initial state ($d^7d^7$).  
When $J_H$ approaches to zero, the above equations reduces to $J=\frac{16 t_3^2}{81 U}$ and $K=-\frac{3}{4}J$ respectively, consistent with Eq.~(7) in Ref. \cite{huimei_prb_2018}. 
When $J_H/U>0.15$, the sign of $J$ changes and grows quickly due to the large $t_3^2$ implying a dominant FM Heisenberg interaction. This precise same path leads to the antiferromagnetic (AFM) Kitaev interaction. However, as we will show below, this weakens the FM  Kitaev interaction from the two-holes $t_{2g}-e_{g}$ exchange paths. The combination of these two results, i.e, incomplete-cancellation of $J$ and almost-cancellation of $K$ is a key reason why some cobaltates do not fall into the Kitaev materials.

Our spin interactions from the $t_2$ and $t_6$ processes are  consistent with the results reported in Ref.  \cite{sano_KHd7_2018}. 
These channels play a minor role in the second-order processes. 
The Kitaev interaction has both FM and AFM contributions from different second order processes. 
The $\Gamma$ interaction only involves two cross terms inside $t_{2g}$ manifolds. The expression of $\Gamma$ and the full expression of $J$ and $K$ from other hopping paths are listed in Table~\ref{table:2nd} in Appendix B. 

\subsection{Two-hole process}
Due to the strong $p-d$  hybridization, contributions from $p$ orbitals are also important. The processes involving $p$ orbitals start at fourth order to introduce coupling between neighbouring Co atoms. The $p$ mediated hopping processes can be included to the effective $d-d$ hoppings.
The leftover fourth-order processes are two-hole and cyclic exchange processes,
which are shown below for completeness. 

\begin{figure}
    \centering
    \includegraphics[width=0.8\columnwidth]{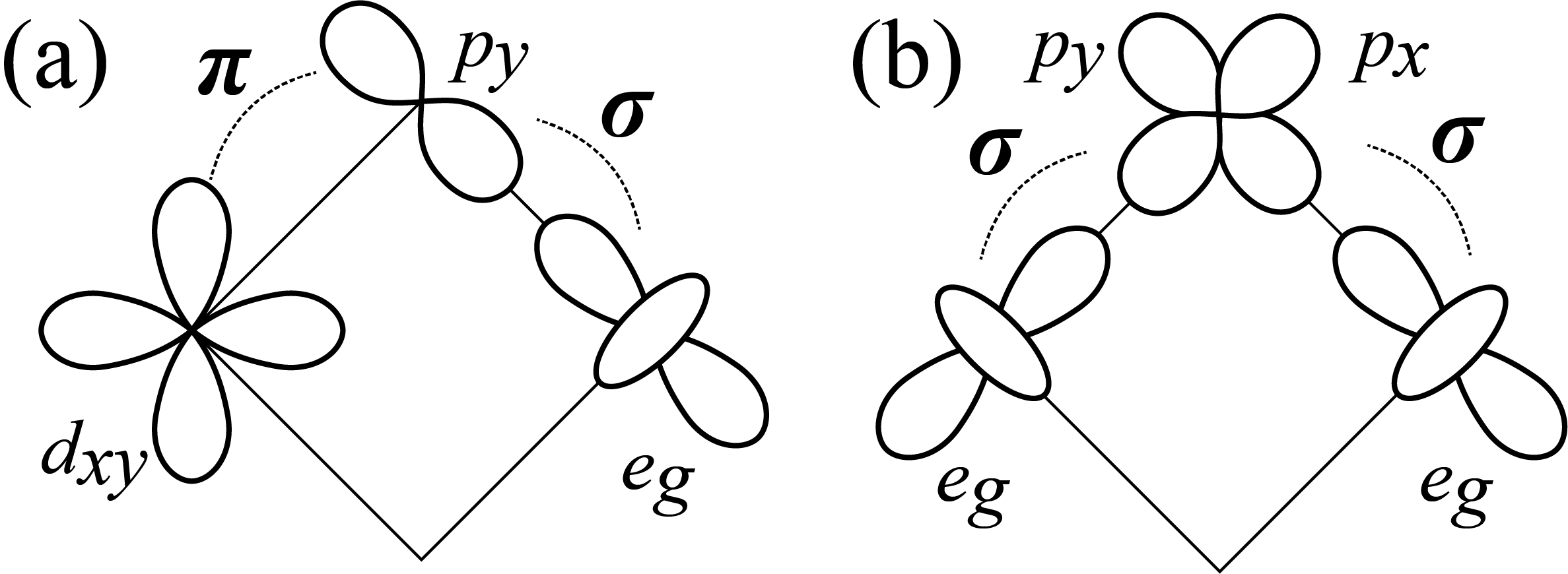}
    \caption{The dominant two-hole processes with two-hole in the same orbital (a) and different orbital (b). }
    \label{fig:B2C2}
\end{figure}

The two-hole processes include intermediate states when the two holes locate at one ligand atom simultaneously. 
The two holes can either locate at the same orbital or different orbitals. The former leads to AFM Heisenberg contributions and the latter prefers FM due to the Hund's coupling of $p$ orbitals $J_{Hp}$. 
The analytical expressions for two-hole processes are listed in Appendix B. 
When expanded to the linear order of $J_{Hp}$, our results are consistent with Ref. \cite{huimei_prb_2018}.
The contributions from $t_{2g}-t_{2g}$ , $t_{2g}-e_g$ and $e_g-e_g$ groups indicated by their hopping integrals  $t_{pd\pi}^4$, $t_{pd\pi}^2 t_{pd\sigma}^2$ and $t_{pd\sigma}^4$, respectively. 
The two dominant two-hole processes are illustrated in Fig.~\ref{fig:B2C2}. Fig.~\ref{fig:B2C2}(a) provides a configuration capable of accommodating two holes at the same orbital. It requires one of the holes coming from the $t_{2g}$ orbitals and the other from the $e_g$ orbitals and the two holes are spin opposite, leading to an AFM Heisenberg term.  It is also worth mentioning that due to the different origins of the holes, the four paths of two-hole processes can be shown in its dependence of $t_{pd\pi}^2 t_{pd\sigma}^2 \left(\dfrac{1}{\Delta_{pd}}+\dfrac{1}{\Delta_{pd}+\Dq}\right)^2$ under perturbation theory. 
Fig.~\ref{fig:B2C2}(b) shows a configuration with two holes at different orbitals. A FM contribution is expected as the two holes prefer align the spins parallel due to Hund's coupling.  Notice that the $e_g-e_g$ processes do not have Kitaev contributions. This is due to the cubic CFS that quenches the orbital angular momentum of the $e_g$ orbitals. 
As analyzed above, the two processes shown in Fig.~\ref{fig:B2C2} have opposite sign and partially cancel with each other. 

\subsection{Cyclic exchange process}
\begin{figure}
    \centering
    \includegraphics[width=0.8\columnwidth]{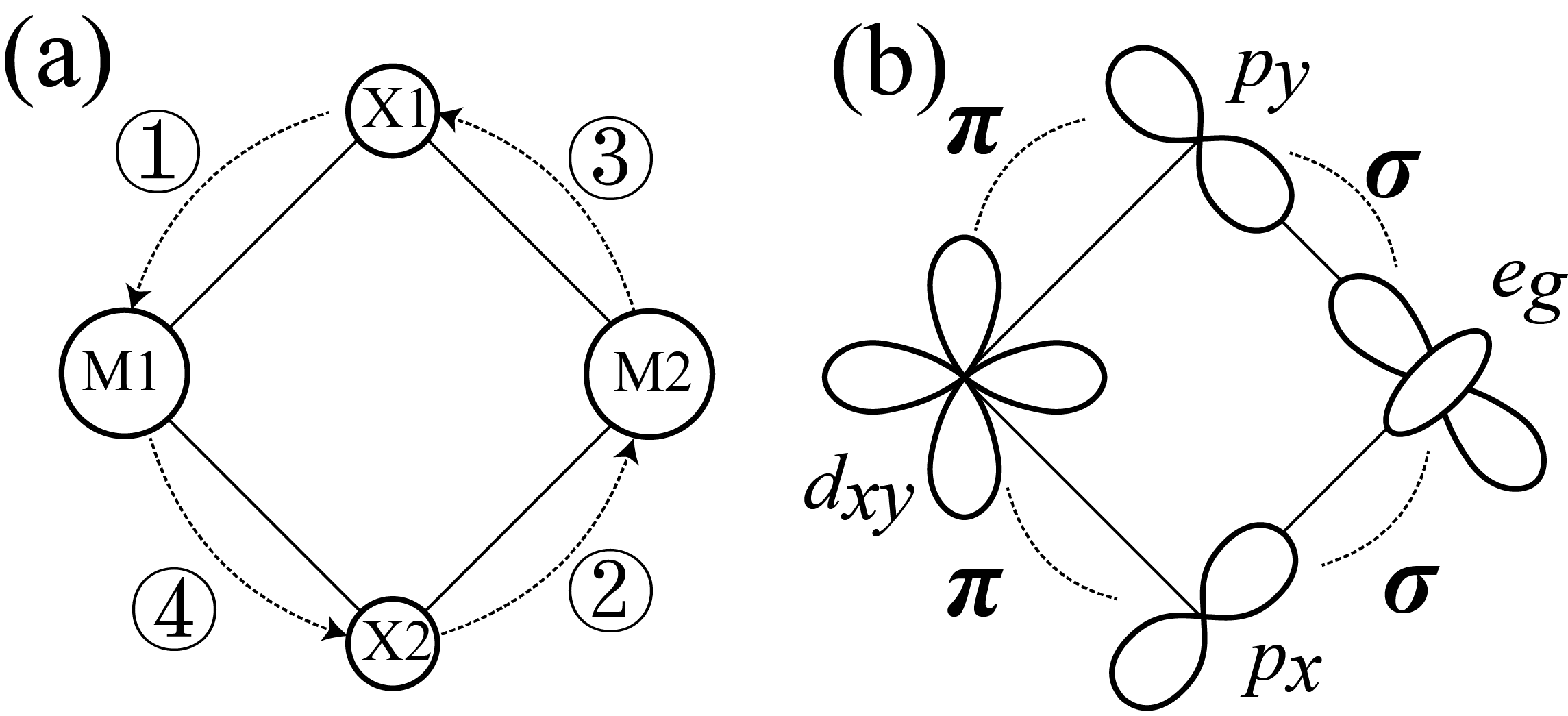}
    \caption{(a) shows the paths for cyclic exchange processes. The related paths are obtained by interchanging \textcircled{1}$\leftrightarrow$\textcircled{2} and/or \textcircled{3}$\leftrightarrow$\textcircled{4}, as well as reversing the cycle direction. (b) The dominant cyclic exchange process.}
    \label{fig:cyclic}
\end{figure}

Another fourth order perturbation process is the cyclic exchange process as shown in Fig~\ref{fig:cyclic}(a). 
It has an intermediate state when each of the two ligand atoms has one hole. Due to the property of identical particles, we cannot distinguish the two holes apart and thus this is a pure quantum mechanical process. 
The dominant process is illustrated in Fig~\ref{fig:cyclic}(b) which involves both $t_{2g}$ and $e_g$ orbitals. The geometry of Fig~\ref{fig:cyclic}(b) requires that the two $\sigma$ processes have to have opposite sign. Thus the cyclic exchange processes are always FM. 

Combining all the significant processes up to fourth order, only one process (Fig.~\ref{fig:B2C2}(a)) can give rise to AFM Heisenberg. Usually this process has comparable magnitude with other paths shown in Fig.~\ref{fig:B2C2}(b) and/or Fig.~\ref{fig:cyclic}(b). Thus the perfect cancellation of Heisenberg cannot be expected generally. 

\section{Application to BCAO}
To apply the above theory, we obtain tight binding parameters for BCAO using DFT calculations. 
The calculation is performed with Vienna \textit{ab initio} Simulation Package (VASP)~\cite{vasp1993} with projector augmented wave (PAW)~\cite{paw1994} potential and Perdew-Burke-Ernzerhof (PBE)~\cite{pbe1996} exchange-correlation functional. 
The cutoff energy of the plane wave basis is set to be 400~eV. $k$-point mesh is $5\times5\times5$. 
The hopping parameters are obtained with  spin-unpolarized calculations without SOC. 
The Wannier90 code~\cite{wannier90_2020} is used to build tight-binding models out of DFT calculation. 
To discuss both 2nd- and 4th-order perturbation process, we build two Wannier models from purely $d$ orbitals and from both $p$ and $d$ orbitals. 
The tight-binding parameters are provided in Appendix A where hoppings in $T_{dd}$ and $T_{dp}$ are read out. 

We confirm that $t_3$ is the largest hopping integral. 
We also note that the indirect hopping between $d_{xz}$ and $d_{yz}$ orbitals through $p$ orbitals, i.e., some part of $t_2$ channels ($t$ in  Ref. \cite{huimei_prb_2018}) cancels with the direct hopping channel and is thus much smaller than $t_3$. 

Using the tight binding parameters and the exchange interactions obtained above, a set of exchange parameters for an idealized honeycomb BCAO are listed in Table~\ref{table:numerics}. 
Here we classify the various second- and fourth-order processes into three groups, depending on whether hoppings are between $t_{2g}-t_{2g}$ orbitals, $t_{2g}-e_g$ orbitals or $e_g-e_g$ orbitals.  They are denoted by A, B, and C, respectively. 1, 2 and 3 represents second-order, two-hole, and cyclic exchange processes respectively, following the same notations used in Ref. \cite{huimei_prb_2018}. Note the the dominant Heisenberg interaction, while almost cancellation of the Kitaev interaction from A1, B2 and B3.

As expected from our analysis, the second-order inter-site $U$ process gives rise to a FM Heisenberg interaction. 
The only significant AFM Heisenberg interaction comes from the two-hole processes (B2) with the two holes at the same orbital. The cancellation of Heisenberg interactions is incomplete.
Tuning the local interaction parameters such as $U$ and $J_H$ can alter the numbers but does not affect the qualitative result that Heisenberg interaction is dominant.

\begin{table}
    \caption{Exchange interactions for the idealized honeycomb BCAO. We set $U=6~\text{eV}$, $J=0.2~U$, $U_p=4~\text{eV}$ $J_{Hp}=1~\text{eV}$ and $\Dq=0.9~\text{eV}$. The SOC for Co atoms is 60~meV. 1, 2 and 3 are inter-site, two-hole and cyclic exchange processes respectively. A, B and C denotes contributions from $t_{2g}-t_{2g}$, $t_{2g}-e_g$ and $e_g-e_g$ processes respectively.  }\label{table:numerics}
    \begin{ruledtabular}
    \begin{tabular}{ccccccccccc}
 & A1 & B1 & C1 & A2 & B2 & C2 & A3 & B3 & C3 & tot \\\hline
$J$ & -5.79 &   0.04 & 0.45 &  0.17 &  14.37 &  -6.3 &  0.1 &  -11.75 &  0 &  -8.7 \\
$K$ &  2.47 &  -0.2 &  0 &  0.71 &  -8.38 &  0 &  -1.03 &  5.87 &  0 &  -0.55 \\
$\Gamma$ & -0.56 &  0 &  0 &  0 &  0 &  0 &  0 &  0 &  0 &  -0.56 \\
    \end{tabular}
    \end{ruledtabular}
\end{table}

\section{trigonal distortion}
The above results are limited to the ideal system where the $J_{\rm eff}=\frac{1}{2}$ picture is intact. 
In materials like BCAO, the trigonal distortion is present, which modifies the ideal $J_{\rm eff}=\frac{1}{2}$ picture. 
In this section, we present how the trigonal distortion alters the spin Hamiltonian and the relative strength of exchange interactions.

The trigonal distortion $H_{\rm trig}=\delta (3 L_Z^2 - 2)$\cite{huimei_prl_2020} breaks the 
$J_{\rm eff}=\frac{1}{2}$,
but 
the lowest states are still doubly degenerate and isolated from other energy states for some ranges of the trigonal-distortion strength $\delta$.\cite{liu2021IJMP,winter_d7_2022}. 
Since $H_{\rm trig}$ commutes with the $Z$ component  of total angular momentum $J_Z$ where $Z=[111]$ in $xyz$ coordinate, see Fig.~1 (a),
$J_Z$ is preserved for a finite $\delta$.
It is useful to write the spin model in the $XYZ$ coordinaite.

The spin Hamiltonian in the global $XYZ$ coordinate is written as
(for $z$-bond)
\begin{equation}\label{eq:XYZ}
    \begin{aligned}
   H_{\rm spin}^z &= \left(
       \begin{array}{ccc}
       J_{XY} + D & E & F   \\
        E & J_{XY} - D &  G \\
        F & G & J_Z\\
    \end{array}
       \right).
    \end{aligned}
\end{equation}
Here we use the notation introduced in Ref. \cite{das_XY_2021}
where $D$ and $F$ corresponds to $A$ and $-\sqrt{2} B$ in Ref. \cite{chaloupka2015PRB}, respectively, which are equivalent to  $J_{ab}$ and $-\sqrt{2} J_{ac}$ in Ref. \cite{Cen2022CP}. 
When the $C_{2v}$ symmetry along the $Y$-axis is intact, $E = G = 0$. 

Rotating the above Hamiltonian to the $xyz$ coordinate, the model is given by the familiar $JK\Gamma\Gamma'$ with a small deviation if the $C_{2v}$ symmetry is broken.
\begin{equation}\label{eq:xyz}
    \begin{aligned}
    H_{\rm spin}^z &= \left(
        \begin{array}{ccc}
        J+\eta & \Gamma & \Gamma_1^\prime   \\
        \Gamma & J-\eta &  \Gamma_2^\prime \\
        \Gamma_1^\prime & \Gamma_2^\prime & J+K\\
    \end{array}
        \right).
    \end{aligned}
\end{equation}

Since the analytical expressions of the exchange interactions including the trigonal distortion are hard to arrive, we present the numerical results which are listed in Table~\ref{table:XYZ} and Table~\ref{table:xyz}. From Table~\ref{table:XYZ} we find the dominant FM XXZ Heisenberg terms ($J_{XY}$ and $J_Z$) as expected. $D$, $E$ and $F$ serve as additional anisotropic terms and are negligible. 
We also find that the contributions from fourth-order processes (process 2 and 3) are non-negligible compared with the inter-site $U$ second order contribution (process 1). 
Converting the coordinate to $xyz$, we find the FM Kitaev are overall negligible as shown in Table~\ref{table:xyz}. Here we ignore the distortion induced hopping for simplicity\cite{stavropoulos_prr_2021,xiaoyu_SIA_2022}, which would not change the main finding. The result is consistent with the analysis from ideal case. Apart from the cancellation of Kitaev term between different processes, the inclusion of the trigonal distortion further weakens the Kitaev term in each processes. 
Since $C_{2v}$ symmetry is intact, $\eta=0$ and $\Gamma_1^\prime = \Gamma_2^\prime$.
Depending on how the layers are stacked, $C_{2v}$ can be broken leading to 
 finite $\eta$ and different $\Gamma^\prime$
 (and a finite $E$ and $G$), which are ignored in this study. 
Other bonds are related with $z$-bond by $C_3$ rotation. 

\begin{table}
    \caption{Exchange interactions (in unit of meV) under the trigonal distortion. 1, 2 and 3 are inter-site, two-hole and cyclic exchange processes respectively. We use the same parameters as shown in the caption of  Table~\ref{table:numerics} with the SOC of Co atom as 60~meV and the trigonal field $\delta=40$~meV. }\label{table:XYZ}
    \begin{ruledtabular}
    \begin{tabular}{ccccc}
 & 1 & 2 & 3 & tot \\\hline
$J_{XY}$ & -6.91 &  7.66 &  -11.88 &  -11.13 \\
$J_Z$ & -1.59 &  2 &  -3.06 &  -2.65\\
$D$ & -0.17 &  0.81 &  -0.28 &  0.35\\
$E$ & 0 &  0 &  0 &  0 \\
$F$ & 0.01 &  -0.55 &  0.55 &  0.02 \\
$G$ & 0 &  0 &  0 &  0 \\
    \end{tabular}
    \end{ruledtabular}
\end{table}

\begin{table}
    \caption{Exchange interactions (in unit of meV) under the trigonal distortion in the $xyz$ basis, converted from Table~\ref{table:XYZ}.}\label{table:xyz}
    \begin{ruledtabular}
    \begin{tabular}{ccccc}
 & 1 & 2 & 3 & tot \\\hline
$J$ & -5.2 &  6.3 &  -9.24 &  -8.14 \\
$K$ &  0.19 &  -1.58 &  0.91 &  -0.47\\
$\eta$ & 0 &  0 &  0 &  0\\
$\Gamma$ & 1.88 &  -2.17 &  2.92 &  2.63 \\
$\Gamma_1^\prime$ & 1.72 &  -1.75 &  2.88 &  2.85 \\
$\Gamma_2^\prime$ & 1.72 &  -1.75 &  2.88 &  2.85 \\
    \end{tabular}
    \end{ruledtabular}
\end{table}

\section{Summary and Discussion}
Kitaev materials refer to materials with the dominant Kitaev interaction over other symmetry-allowed interactions. 
Recently it was suggested that $3d^7$ honeycomb cobaltates are Kitaev candidates \cite{huimei_prb_2018,huimei_prl_2020,sano_KHd7_2018} which has extended our search for the KSL.
However, the application of the proposal was questioned by other theoretical and experimental works on BCAO and BCPO. \cite{das_XY_2021,maksimov_ab_2022,winter_d7_2022,halloran_geometrical_2022}
We investigate a possible origin of the two different proposals, non-Kitaev vs. Kitaev cobaltates.

The original proposal made by
Liu et al and Sano et al\cite{huimei_prb_2018,sano_KHd7_2018} is based on the idea
that AFM Heisenberg interaction from the $t_{2g}-e_g$ and FM Heisenberg interaction from the $e_g-e_g$ paths almost cancel out each other, while the $t_{2g}-e_g$ channel generates FM Kitaev interaction, resulting in the dominant Kitaev interaction. These theories assumed the negligible intraorbital $t_3$ hopping integral.
We expect that $t_3$ is the largest hopping integral, and show that the exchange path associated with $t_3$
turns the story around. The $t_3$ exchange channel generates FM Heisenberg and AFM Kitaev interactions. Combining with other contributions, the FM Heisenberg interaction from the $t_3$ path becomes the dominant interaction, while
the Kitaev interaction from the $t_3$ and the $t_{2g}-e_g$ paths almost cancel out each other. 
Applying our theory to BCAO, we find that $t_3$ is indeed largest consistent with \cite{maksimov_ab_2022}, which seems to be the case for other cobaltates \cite{winter_d7_2022}.

However, in NSCO, the intersite $t_{2g}-e_g$ (denoted by $t_6$) hopping increases while $t_3$ decreases even though it is still the largest  hopping.\cite{Kim2022JPCM} In this case, the AFM Kitaev interaction from the $t_3$ channel becomes weaker
while the $t_6$ channel boosts the FM Kitaev interaction. The almost-cancellation of the Kitaev interaction discussed above no longer occurs. At the same time, the Heisenberg interaction from the $t_3$ channel becomes weaker. All together, it implies that the Kitaev interaction is comparable to or even larger than the reduced Heisenberg interaction.

Under the trigonal distortion, the doublet is modified from the $J_{\rm eff}=\frac{1}{2}$.
This leads to additional anisotropic spin interactions. Experimentally, BCAO exhibits strong anisotropic g-factors\cite{das_XY_2021}, implying that the trigonal distortion is relatively strong compared with SOC for Co atom. 
We find that the Heisenberg interaction becomes the XXZ type when the trigonal distortion is introduced, which is rather expected, since the ideal limit has the dominant isotropic Heisenberg interaction. 
We did not consider the third n.n. Heisenberg interaction $J_3$, as our motivation is to find the origin of the debate over the dominant n.n. Kitaev vs. Heisenberg interactions. The importance of $J_3$ can be found in Ref.  \cite{Nair2018PRB,maksimov_ab_2022,winter_d7_2022,halloran_geometrical_2022}.

While a material-dependent analysis is required to take into the local environment such as bond lengths and bond angles, 
we expect that the dominant FM XXZ interaction is common in
BCAO and BCPO due to the large intraorbital hopping integral $t_3$ and Hund's coupling.
A target Kitaev cobaltate displaying the KSL can be engineered with effectively reduced intraorbital $t_3$ and enhanced $p$ orbital mediated hoppings, which will move the system towards the dominant Kitaev regime, as proposed in Ref. \cite{huimei_prb_2018,sano_KHd7_2018}. 
Future theoretical and experimental works are needed to discover Kitaev honeycomb cobaltates.

\section*{Acknowledgement}
We thank D. Churchill, Y. B. Kim, and H. Liu for useful discussions. We also thank K. Ross bringing BCAO and BCPO to our attention.
This work is supported by the Natural Sciences and Engineering Research Council of Canada (NSERC) and the Center for Quantum Materials at the University of Toronto. H.Y.K acknowledges the support by the Canadian Institute for Advanced Research (CIFAR) and the Canada Research Chairs Program. Computations were performed on the Niagara supercomputer at the SciNet HPC Consortium. SciNet is funded by: the Canada Foundation for Innovation under the auspices of Compute Canada; the Government of Ontario; Ontario Research Fund - Research Excellence; and the University of Toronto.

\bibliography{references}

\appendix
\onecolumngrid
\section{tight-binding parameters}
We construct two tight-binding model for BCAO. One includes only $d$ orbitals of Co atom. There are in total 10 $d$ orbitals in one unit cell. The onsite Hamiltonian as well as the hoppings between $d-d$ orbitals obtained from DFT and Wannier calculations are given below:
\begin{equation*}
       H_d^{10}= \left(
    \begin{array}{ccccc}
4850.77 & 0 & 7.75 & 8.45 & -16.2 \\
0 & 4850.77 & 14.24 & -13.83 & -0.4 \\
7.75 & 14.24 & 3952.74 & 38.87 & 38.87 \\
8.45 & -13.83 & 38.87 & 3952.74 & 38.87 \\
-16.2 & -0.4 & 38.87 & 38.87 & 3952.74 \\
    \end{array}\right),
\end{equation*}
\begin{equation*}
       T_{dd}^{10}= \left(
    \begin{array}{ccccc}
-40.89 & 0.4 & -17.87 & 34.01 & -14.28 \\
0.4 & -37.36 & -0.43 & 14.32 & 45.74 \\
-17.87 & -0.43 & 66.18 & -19.62 & 35.26 \\
34.01 & 14.32 & -19.62 & 66.49 & 26.71 \\
-14.28 & 45.74 & 35.26 & 26.71 & -295.49 \\
    \end{array}\right).
\end{equation*}
Another tight-binding model includes both Co $d$ atoms and O $p$ orbitals. There are in total 34 orbitals. The onsite Hamiltonian for both $d$ orbitals and $p$ orbitals, as well as the direct hopping between $d-d$ and the hopping between $p-d$ are given as:
\begin{equation*}
       H_{d}^{34}= \left(
    \begin{array}{ccccc}
3721.77 & 0 & 1.01 & 3.83 & -4.83 \\
0 & 3721.77 & 5 & -3.37 & -1.63 \\
1.01 & 5 & 3543.68 & 10.93 & 10.93 \\
3.83 & -3.37 & 10.93 & 3543.68 & 10.93 \\
-4.83 & -1.63 & 10.93 & 10.93 & 3543.68 \\
    \end{array}\right),
\end{equation*}
\begin{equation*}
       T_{dd}^{34}= \left(
    \begin{array}{ccccc}
-28.31 & -9.34 & -5.14 & 5.08 & -22.57 \\
-9.34 & -45.42 & 12.23 & 0.92 & -104.37 \\
-5.14 & 12.23 & 82.14 & -108.02 & 14.97 \\
5.08 & 0.92 & -108.02 & 64.46 & -9.19 \\
-22.57 & -104.37 & 14.97 & -9.19 & -249.37 \\
    \end{array}\right),
\end{equation*}
\begin{equation*}
       H_{p}^{34}= \left(
    \begin{array}{ccc}
43.28 & -890.74 & -890.74 \\
-890.74 & 43.28 & -890.74 \\
-890.74 & -890.74 & 43.28 \\
    \end{array}\right),
\end{equation*}
\begin{equation*}
       T_{dp}^{34}= \left(
    \begin{array}{ccc}
-1151.92 & -61.22 & 175.07 \\
653.14 & 36.17 & -133.84 \\
5.62 & -40.96 & 1.85 \\
163.59 & 32.32 & 625.23 \\
11.17 & 691.32 & 28.09 \\

    \end{array}\right).
\end{equation*}

\section{$J$, $K$, and $\Gamma$ interactions for ideal octahedra cage}
For the Heisenberg and Kitaev interactions from the intersite exchange paths are listed in Table~\ref{table:2nd}. There are several paths and a detailed balance among them may change the order of the dominant interactions.
On the other hand for the $\Gamma$ interaction from the intersite processes, there are only two contributions from $t_1 t_2$ and
$t_2 t_3$. 



\begin{table*}
\caption{Intersite $U$ Exchange Interactions for ideal case.} \label{table:2nd}
\begin{ruledtabular}
\begin{tabular}{c|cc}
 & $J$ & $K$   \\\hline\\[-5pt]
 $t_1^2$ & $\dfrac{1}{486} \biggl(-\dfrac{171}{U-3\JH}+\dfrac{259}{U+\JH}+\dfrac{44}{U+4\JH}\biggr)$&
 $\dfrac{1}{243}\biggl(\dfrac{45}{U-3\JH}+\dfrac{11}{U+\JH}+\dfrac{28}{U+4\JH}\biggr)$  \\[10pt]
 $t_2^2$  & $\dfrac{1}{54} \biggl(-\dfrac{21}{U-3\JH}+\dfrac{29}{U+\JH}+\dfrac{4}{U+4\JH}\biggr)$ &
 $\dfrac{1}{243}\biggl(-\dfrac{81}{U-3\JH}+\dfrac{73}{U+\JH}-\dfrac{4}{U+4\JH}\biggr)$  \\[10pt]
 $t_3^2$  & $\dfrac{2}{243}\biggl(-\dfrac{27}{U-3\JH}+\dfrac{43}{U+\JH}+\dfrac{8}{U+4\JH}\biggr)$ & 
 $\dfrac{2}{81}\biggl(\dfrac{3}{U-3\JH}-\dfrac{7}{U+\JH}-\dfrac{2}{U+4\JH}\biggr)$   \\[10pt]
 $t_4^2$ & $\dfrac{100}{81(U+2\JH)}$ & 0   \\[10pt]
 $t_5^2$ & $\dfrac{100}{81(U+2\JH)}$ & 0   \\[10pt]
 $t_6^2$ & 
 \parbox[t]{6cm}{
 $\dfrac{5}{243} \biggl( -\dfrac{27}{U-3\JH+\Dq} + \dfrac{43}{U+\JH+\Dq}$ \\ 
 $ + \dfrac{8}{U+4\JH+\Dq} + \dfrac{24}{U+2\JH-\Dq} \biggr)$ } 
 &
 \parbox[t]{6cm}{
 $\dfrac{5}{243} \biggl( -\dfrac{9}{U - 3\JH + \Dq} + \dfrac{1}{U + \JH + \Dq}$ \\ 
 $ - \dfrac{4}{U + 4\JH + \Dq} - \dfrac{12}{U + 2\JH - \Dq} \biggr)$ } \\[30pt]
 $t_1 t_3$ & $\dfrac{4}{243}\biggl(\dfrac{18}{U-3\JH}-\dfrac{8}{U+\JH}+\dfrac{5}{U+4\JH}\biggr)$ &
 $\dfrac{1}{243}\biggl(-\dfrac{63}{U-3\JH}+\dfrac{31}{U+\JH}-\dfrac{16}{U+4\JH}\biggr)$  \\[10pt]
 \hline  
 \multicolumn{3}{c}{} \\[-5pt]
 \multicolumn{3}{c}{$    \Gamma = \dfrac{4 t_1 t_2 }{81}\left(\dfrac{3}{U-3\JH}-\dfrac{7}{U+\JH}-\dfrac{2}{U+4\JH}\right) 
    + \dfrac{t_2 t_3}{243}\left(-\dfrac{63}{U-3\JH}+\dfrac{31}{U+\JH}-\dfrac{16}{U+4\JH}\right)$}\\[10pt]
\end{tabular}
\end{ruledtabular}
\end{table*}

The two hole (denoted by superscript $2h$) and cyclic exchange (superscript $cyclic$) processes are shown below. 

\begin{eqnarray*}
J^{2h} &=& \biggl[-\dfrac{80}{81} \dfrac{1}{ \left(2\Delta_{pd}+U_p-3J_{Hp}\right)} 
+\dfrac{304}{243} \dfrac{1}{ \left(2\Delta_{pd}+U_p-J_{Hp}\right)}
+ \dfrac{32}{243} \dfrac{1}{ \left(2\Delta_{pd}+U_p+2J_{Hp}\right)}\biggr] 
\times t_{pd\pi}^4 \left(\dfrac{1}{\Delta_{pd}^2}\right)\\
& & + \biggl[-\dfrac{10}{27} \dfrac{1}{2\Delta_{pd}+U_p-3J_{Hp}+\Dq} 
+ \dfrac{250}{243} \dfrac{1}{\left(2\Delta_{pd}+U_p-J_{Hp}+\Dq\right)}
+\dfrac{80}{243}\dfrac{1}{\left(2\Delta_{pd}+U_p+2J_{Hp}+\Dq\right)^2}
 \biggr] \\
 & &\times
 t_{pd\pi}^2 t_{pd\sigma}^2
\left(\dfrac{1}{\Delta_{pd}}+\dfrac{1}{\Delta_{pd}+\Dq}\right)^2\\
& &+ \biggl[-\dfrac{200}{81} \dfrac{1}{\left(2\Delta_{pd}+U_p-3J_{Hp}+2\Dq\right)}
+\dfrac{200}{81} \dfrac{1}{\left(2\Delta_{pd}+U_p-J_{Hp}+2\Dq \right)}
\biggr] 
\times t_{pd\sigma}^4\left(\dfrac{1}{ \Delta_{pd} + \Dq }\right)^2,
\end{eqnarray*}

\begin{eqnarray*}
K^{2h} & = & \biggl[\dfrac{40}{81} \dfrac{1}{ \left(2\Delta_{pd}+U_p-3J_{Hp}\right)} 
-\dfrac{56}{243} \dfrac{1}{ (2\Delta_{pd}+U_p-J_{Hp})}
+\dfrac{32}{243} \dfrac{1}{ \left(2\Delta_{pd}+U_p+2J_{Hp}\right)}\biggr]
\times t_{pd\pi}^4 \left(\dfrac{1}{\Delta_{pd}^2}\right)\\
& &+\biggl[-\dfrac{10}{81} \dfrac{1}{2\Delta_{pd}+U_p-3J_{Hp}+\Dq} 
- \dfrac{50}{243} \dfrac{1}{\left(2\Delta_{pd}+U_p-J_{Hp}+\Dq\right)}
- \dfrac{40}{243} \dfrac{1}{\left(2\Delta_{pd}+U_p+2J_{Hp}+\Dq\right)}\biggr]\\
& & \times t_{pd\pi}^2 t_{pd\sigma}^2
\left(\dfrac{1}{\Delta_{pd}}+\dfrac{1}{\Delta_{pd}+\Dq}\right)^2,\\
\end{eqnarray*}

\begin{eqnarray*}
J^{cyclic} & = &\left[\dfrac{2}{81\Delta_{pd}}\right]t_{pd\pi}^4 \left(\dfrac{1}{\Delta_{pd}^2}\right)
+ \left[-\dfrac{40}{81}\dfrac{1}{2\Delta_{pd} +  \Dq }\right]t_{pd\pi}^2 t_{pd\sigma}^2
    \left(\dfrac{1}{\Delta_{pd}}+\dfrac{1}{\Delta_{pd}+\Dq}\right)^2,\\
\end{eqnarray*}

\begin{eqnarray*}
K^{cyclic} & = & \left[-\dfrac{20}{81\Delta_{pd}}\right]t_{pd\pi}^4 \left(\dfrac{1}{\Delta_{pd}^2}\right)
+ \left[\dfrac{20}{81}\dfrac{1}{2\Delta_{pd} +  \Dq }\right]t_{pd\pi}^2 t_{pd\sigma}^2
    \left(\dfrac{1}{\Delta_{pd}}+\dfrac{1}{\Delta_{pd}+\Dq}\right)^2.
\end{eqnarray*}

\nocite{*}
\end{document}